\documentclass[11pt,letterpaper]{article}%
\usepackage{setspace}
\usepackage{color}
\usepackage{amsmath}
\usepackage{amsfonts}
\usepackage{verbatim}
\usepackage{amssymb}
\usepackage{graphicx}
\usepackage{color}%
\providecommand{\U}[1]{\protect\rule{.1in}{.1in}}
\definecolor{dgreen}{rgb}{0.,0.6,0.}
\definecolor{olive}{rgb}{0.3, 0.4, .1}
\begin{document}

\title{Gribov ambiguity and degenerate systems}
\author{Fabrizio Canfora$^{1,2}$, Fiorenza de Micheli$^{1,3}$, Patricio
Salgado-Rebolledo$^{1,4,5}$ and Jorge Zanelli$^{1,2}$\\$^{1}$ {\small \textit{Centro de Estudios Cient\'{\i}ficos (CECS), Arturo Prat
514, Valdivia, Chile.}}\\$^{2}${\small \textit{Universidad Andr\'{e}s Bello, Av. Rep\'{u}blica 440,
Santiago, Chile.}}\\$^{3}${\small \textit{Instituto de F\'{\i}sica, Pontificia Universidad
Cat\'{o}lica de Valpara\'{\i}so, Casilla 4059, Valpara\'{\i}so, Chile.}}\\$^{4}${\small \textit{Departamento de F\'{\i}sica, Universidad de
Concepci\'{o}n, Casilla 160-C, Concepci\'{o}n, Chile.}}\\$^{5}${\small \textit{Physique Th\'{e}orique et Math\'{e}matique,
Universit\'{e} Libre de Bruxelles and}}\\{\small \textit{International Solvay Institutes, Campus Plaine C.P. 231,
B-1050 Bruxelles, Belgium.}}\\{\small \texttt{canfora@cecs.cl, zuccavuota@gmail.com, pasalgado@udec.cl,
z@cecs.cl}}}
\maketitle

\begin{abstract}
The relation between Gribov ambiguity and degeneracies in the symplectic
structure of physical systems is analyzed. It is shown that, in
finite-dimensional systems, the presence of Gribov ambiguities in regular
constrained systems (those where the constraints are functionally
independent) always leads to a degenerate symplectic structure upon Dirac
reduction. The implications for the Gribov-Zwanziger approach to QCD are discussed.

\end{abstract}

\newpage


\section{Introduction}



\subsection{Gribov problem and the Zwanzinger restriction}

In his seminal paper, Gribov showed that a standard gauge condition, such as
the Coulomb or the Landau choices, fail to provide \textit{proper gauge
fixings}\footnote{A gauge fixing is called proper if it intersects all gauge
orbits only once \cite{Dirac,HT}.} in Yang-Mills theories \cite{Gri78}. This
so-called Gribov problem, that affects non-abelian gauge theories, means that
a generic gauge fixing intersects the same gauge orbit more than once (Gribov
copies) and may fail to intersect others. Algebraic gauge conditions free of
Gribov ambiguities are possible, but those choices are affected by severe
technical problems as, for instance, incompatibility with the boundary
conditions that must be imposed on the gauge fields in order to properly
define the configuration space for the theory \cite{DeW03}. Additionally,
Singer \cite{Singer} showed that Gribov ambiguities occur for all gauge fixing
conditions involving derivatives (see also \cite{Jackiw}), and moreover, the
presence of the Gribov problem breaks BRST symmetry at a non-perturbative
level \cite{Fuj}\footnote{See also \cite{Lavrov,Lavrov2}.}.

The Gribov problem occurs because it is generically impossible to ensure
positive definiteness of the Faddeev-Popov (\textbf{FP}) determinant
everywhere in functional space. The configurations for which the the FP
operator develops a nontrivial zero mode are those where the gauge condition
becomes `tangent' to the gauge orbits and it therefore fails to intersect
them. The Gribov horizon (\textbf{GH}), where this happens, marks the boundary
beyond which the gauge condition intersects the gauge orbits more than once
(Gribov copies). The appearance of Gribov copies invalidates the usual
approach to the path integral and one way to avoid overcounting is to restrict
the sum over field configurations to the so-called Gribov region around
$A_{\mu}=0$, where the FP operator is positive definite (see, in particular,
\cite{Gri78,Zw82,Zw89,DZ89,Zwa96,Va92}).

In the case of flat topologically trivial space-time, the restriction to the
Gribov region coincides with the usual perturbation theory around $A_{\mu}=0$
(with respect to a suitable functional norm \cite{Va92}). The restriction to
the first Gribov region takes into account the infrared effects related to the
\textit{partial} elimination of the Gribov copies, in the sense that it only
guarantees the exclusion of those copies obtained by (\textquotedblleft
small") gauge transformations perturbatively connected to $A_{\mu}=0$
\cite{Zw89,MaggS,Gracey}. It has been shown that non-perturbatively accessible
gauge copies may exist within the Gribov region as well if the space-time is
not flat or is topologically nontrivial \cite{new1,new2,new4}. The complete
elimination of the gauge copies if the space is not flat or its topology is
nontrivial can be a very difficult issue and here we do not consider those
possibilities, restricting the path-integral to the first Gribov region.

Remarkably enough, the partial elimination of Gribov copies in perturbation
theory is related to the non-perturbative infrared physics \cite{Gri78}. The
non-perturbative input in the modified path-integral is the restriction to the
Gribov region. When one takes into account the presence of suitable
condensates \cite{SoVae2,SoVar,SoVar2,SoVar3,soreprl} the agreement with
lattice data is excellent \cite{DOV,CucM}. Moreover, within this approach, it
has been possible to solve an old issue on the Casimir energy and force for
the Yang-Mills field in the MIT bag model \cite{CasimirRGZ}.

A common criticism to the Gribov-Zwanziger approach that restricts the
functional space to the Gribov region is that it goes against the Feynman's
postulate of summing over all histories. There are various arguments that
answer this criticism. First of all, the configurations outside the Gribov
region are copies of some configuration inside the Gribov region \cite{Zw89}
\cite{DZ89}. Therefore the Gribov restriction avoids overcounting and no
relevant physical configurations are lost. Second and most importantly, this
framework considerably improves the analytic results of standard perturbation
theory like the glueball spectrum, which closely reproduces the lattice
results \cite{soreprl}.

Hence, it is natural to look for examples in a context simpler than Yang-Mills
theory, in which the issue of the Gribov restriction can be directly analyzed.
Here we show in some toy models with finite number of degrees of freedom that
it may not be necessary to impose the Gribov restriction ``by hand" but it
arises naturally from the dynamics of the system.


\subsection{Gribov problem and dynamic degeneracy}

In Dirac's formalism for constrained systems \cite{Dirac,HT} gauge-invariant
mechanical systems are characterized by the presence of first class
constraints $G_{i}\approx0$, $i=1,\cdots,n$. Gauge fixing in those systems is
achieved by the introduction of $n$ gauge conditions $\phi_{i}\approx0$, so
that the $2n$ constraints $\{G_{i},\phi_{j}\}$ become a second class set. In
this context, the Gribov problem is the statement that the second class nature
of this set cannot hold globally: the Dirac matrix defined by their Poisson
brackets is not invertible everywhere in phase space, it is
\textit{degenerate}.

Degenerate Hamiltonian systems on the other hand, are those whose symplectic
form is not invertible in a subset $\Sigma$ of phase space $\Gamma$
\cite{STZ}. In classical degenerate systems the evolution takes place over
non-overlapping causally disconnected subregions of the phase space separated
by degenerate surfaces $\Sigma$. This means that if a system is prepared in
one subregion, never evolves to a state in a different subregion. This still
holds in the quantum domain for some simple degenerate systems \cite{dMZ}.
Degenerate systems are ubiquitous in many areas of physics, from fluid
dynamics \cite{REF-2} to gravity theories in higher dimensions
\cite{example1,example2}, in the strong electromagnetic fields of quasars
\cite{Brennan-Gralla-Jacobson}, and in systems such as massive bi-gravity
theory \cite{BHHMT}, which has been shown to possess degenerate sectors where
the degrees of freedom change from one region of phase space to another
\cite{Banados-et-al}.

Here we will show that Gribov ambiguity and the existence of degeneracies are
related problems, and that the GH can be identified as a surface of degeneracy
$\Sigma$. This means that the system would be naturally confined to a region
surrounded by a horizon, exactly as proposed by Zwanziger \cite{Zw82}. This
interpretation of the GH as surface of degeneracy that acts as a boundary
beyond which the evolution cannot reach, makes the restriction in the sum over
histories a natural prescription and not ad-hoc one.


\section{Degenerate systems}

\label{sec_deg_syst}
We now briefly review classical \cite{STZ} and quantum \cite{dMZ} degenerate
systems. In order to fix ideas, let's consider a system described by the first
order action,
\begin{equation}
I[u] =\int\,\mathrm{d} t \left(  X_{A}(u)\dot{u}^{A} - H(u)\right)  \, ,\,
\mbox{with  } A=1,\ldots, N\, . \label{first_order_action}%
\end{equation}
This action can be interpreted in two not exactly equivalent ways: A) The
$u^{A}$'s are $N$ generalized coordinates and $L(u, \dot{u}) = X(u)_{A}\dot
{u}^{A} - H(u)$ is the Lagrangian; B) The $u^{A}$'s are non-canonical
coordinates in a $N$-dimensional phase space $\Gamma$, where $N$ is
necessarily even and (\ref{first_order_action}) gives the action in
Hamiltonian form.

In the first approach, for each $u$ there is a canonically conjugate momentum
a the $2N$-dimensional canonical phase space $\tilde{\Gamma}$ given by
\begin{equation}
p_{A}=\frac{\partial L}{\partial\dot{u}^{A}}\,.
\end{equation}
In this case, this definition gives a set of primary constraints,
\begin{equation}
\Phi_{A}=p_{A}-X_{A}(u)\approx0\,,
\end{equation}
whose (canonical) Poisson brackets define the antisymmetric matrix
\begin{equation}
\lbrack\Phi_{A},\Phi_{B}]=\partial_{A}X_{B}-\partial_{B}X_{A}\equiv\Omega
_{AB}(u)\,.
\end{equation}
If $\Omega_{AB}$ is invertible --which requires $N$ to be even--, the
constraints $\Phi_{A}\approx0$ are second class and $\Omega_{AB}(u)$ gives the
Dirac bracket necessary to eliminate them.\footnote{Since $\Omega_{AB}$ is a
curl, it satisfies the identity $\partial_{A}\Omega_{BC}+\partial_{B}%
\Omega_{CA}+\partial_{C}\Omega_{AB}=0$ (or, $\Omega=dX\Longrightarrow
d\Omega\equiv0$).} Elimination of these second class constraints in the
$2N$-dimensional canonical phase space $\tilde{\Gamma}=\{u^{A},p_{A}\}$
corresponds to choosing half of the $u$'s as coordinates and the rest as
momenta, and $\Omega_{AB}(u)$ will be identified as the (not necessarily
canonical) pre-symplectic form in the reduced $N$-dimensional phase space
$\Gamma$. In fact, in the Hamiltonian approach the pre-symplectic form can be
read from the equations of motion for the action (\ref{first_order_action}),
\begin{equation}
\Omega_{AB}(u)\,\dot{u}^{A}+E_{A}(u)=0\,, \label{eq.motion}%
\end{equation}
where
\begin{equation}
\Omega_{AB}\equiv\partial_{A}X_{B}(u)-\partial_{B}X_{A}(u)\,,\quad
\mbox{and}\quad E_{A}\equiv\partial_{A}H(u)\,.
\end{equation}
This reasoning shows that in the open sets where $\Omega_{AB}$ is invertible,
the Lagrangian and Hamiltonian versions of this system are equivalent. In this
case the inverse symplectic form, $\Omega^{AB}$, defines the Poisson bracket
for the theory in (not necessarily canonical) coordinates
\begin{equation}
\Omega^{AB}=[u^{A},u^{B}]\,. \label{cb}%
\end{equation}

In what follows, we will refer to $\Gamma$ as the phase space where $u$ are
the coordinates. The pre-symplectic form $\Omega_{AB}(u)$ is a function of the
phase space coordinates $u^{A}$ and its determinant can vanish on some subset
$\Sigma\subset\Gamma$ of measure zero. Degenerate systems are characterized by
having a pre-symplectic form whose rank is not constant throughout phase
space. Moreover, in its evolution a degenerate system can reach a degenerate
surface $\Sigma$ where $\det[\Omega_{AB}]=0$ in a finite time,
\begin{equation}
\Sigma=\{u\in\Gamma\,|\,\Upsilon(u)=0\}\,, \label{degeneracy}%
\end{equation}
where $\Upsilon(u)=\epsilon^{A_{1}A_{2}\cdots A_{N}}\Omega_{A_{1}A_{2}}%
\cdots\Omega_{A_{N-1}A_{N}}$ is the Pfaffian of $\Omega_{AB}$, and
$\det[\Omega_{AB}]=(\Upsilon)^{2}$.

Generically, a degenerate surface represents a co-dimension one submanifold in
phase space and, as shown in \cite{STZ}, the classical evolution cannot take
the system across $\Sigma$. The equations of motion (\ref{eq.motion}) can be
solved for $\dot{u}^{A}$ provided $\Omega_{AB}$ can be inverted. Moreover, the
velocity diverges in the vicinity of $\Sigma$, and if $\Delta(u)|_{\Sigma}=0$
is a simple zero, the velocity changes sign across $\Sigma$. Therefore an
initial state on one side of $\Sigma$ could never reach the other: there is no
causal connection between configurations on opposite sides of $\Sigma$. This
degeneracy surface $\Sigma$ acts as a source or sink of orbits, splitting the
phase space into causally disconnected, non overlapping regions.

\begin{figure}[ptbh]
\centering
\includegraphics[width=7cm]{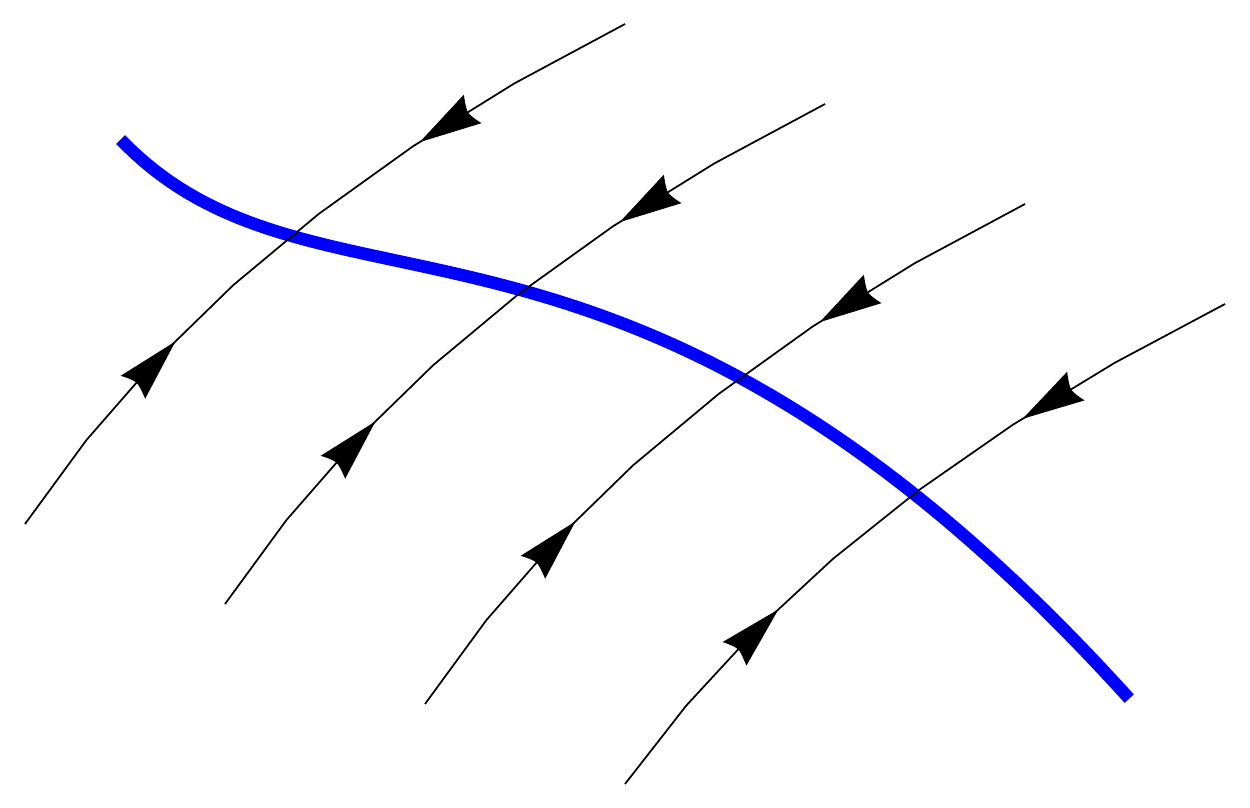} \qquad\qquad
\includegraphics[width=7cm]{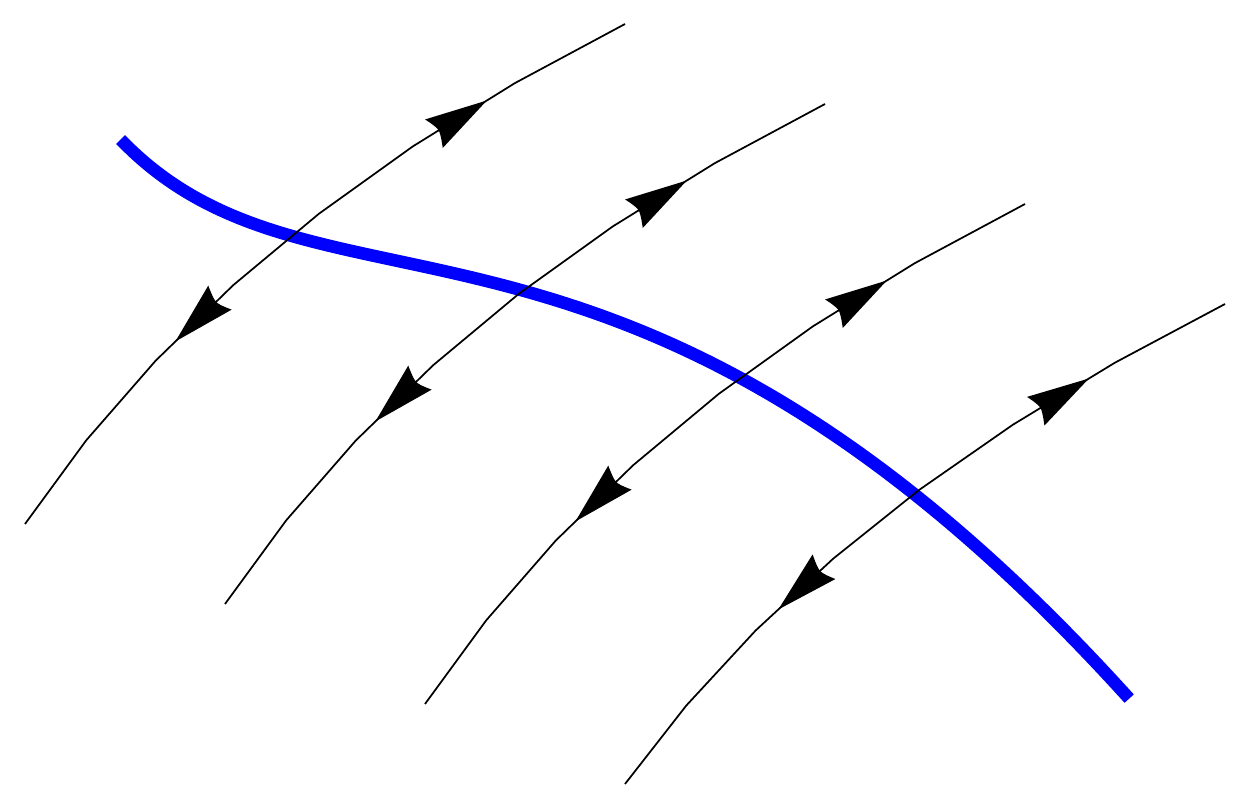}\caption{ Qualitative flow of the orbits
near the degeneracy surface (blue lines), which can act as a sink or as a
source.}%
\label{fig_degeneracy_qualitative_flow}%
\end{figure}

In the quantum case, the degeneracy of the symplectic form becomes the
singular set of the Hamiltonian and the corresponding Hilbert space
$\mathcal{H}$ is endowed with a weighted scalar product,
\begin{equation}
<\psi_{1},\psi_{2}>=\int\mathrm{d}V\psi_{1}^{\ast}\,w\,\psi_{2}\,\;,
\label{scalar_product}%
\end{equation}
where $\mathrm{d}V=\sqrt{g}\,\mathrm{d}^{n}u$ is the volume form and the
weight $w(u)$ is the Pfaffian $\Upsilon(u)$ of the symplectic form
$\Omega_{AB}$
\begin{equation}
w(u):=\sqrt{\det\left[  \Omega_{AB}(u)\right]  }=\Upsilon(u)\,,
\end{equation}
defined in order for the Hamiltonian to be symmetric and for the norm in
$\mathcal{H}$ to be positive definite.

Since singular points must be excluded from the domain of the Hamiltonian
operator, for consistency they should also be excluded from the domain of the
wave functions. This means that the Hilbert space includes wave functions that
can be discontinuous at the degenerate surfaces. Allowing discontinuous wave
functions implies that the solutions can have support restricted to a single
region bounded by $\Sigma$. Therefore the Hilbert space is a direct sum of
orthogonal subspaces of functions defined on each side of the degenerate
surface and, in complete analogy with the classical picture, there is no
quantum tunneling across $\Sigma$.


\section{Gauge Fixing and Gribov ambiguity}

\label{gf}

The quantum description of a gauge-invariant system can be achieved by first
fixing the gauge and then applying the quantization prescription to the
remaining classical degrees of freedom. Let $\Gamma$ be a phase space
described by generalized coordinates $u^{A}$ ($A=1,2,\cdots,N$), endowed with
a symplectic form $\Omega_{AB}(u)$ everywhere invertible. Consider now an open
patch of the phase space $\Gamma$ where a system has local symmetries
generated by a set of first class constraints $\phi_{i}(u)\approx0$,
$(i=1,\ldots,n<N/2)$. Following Dirac's procedure, for a system with $n$ first
class constraints, an equal number of gauge fixing conditions,
\begin{equation}
G_{i}(u)\approx0\,,\,\;\quad i=1,\ldots,n\;, \label{gc}
\end{equation}
must be included so that the whole set of constraints
\begin{equation}
\{\gamma_{I}\}=\{G_{i},\phi_{j}\}\,,\,\;\quad I=1,\ldots,2n<N\,, \label{set}%
\end{equation}
is second class (see \cite{Dirac}). In order to define a proper gauge fixing,
two conditions must be fulfilled: every orbit must intersect the surface
defined by the set $\{G_{i}\}$ in $\Gamma$ (\textit{accessibility}), and orbits can't
intersect the surface defined by $\{G_{i}\}$ more than once (\textit{complete gauge
fixation}) \cite{HT}. In other words, the surface in phase space defined by
the gauge conditions (\ref{gc}) must intersect every orbit once and only once.

\begin{figure}[th]
\begin{minipage}[b]{17cm}
\begin{center}
\hspace{1.0cm}
\includegraphics[scale=3.5]{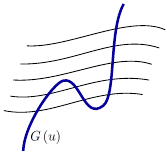}
\caption{The gauge condition $G_{i}(u)\approx 0$ (thick line) intersects the gauge orbits (thin lines) more than once provided there exist points where the orbits run tangent to the gauge condition.}
\end{center}
\end{minipage} \label{fig_gauge_orbits}\end{figure}

The submanifold defined by setting the constraints $\{\gamma_{I}\}$ strongly equal
to zero, corresponds to the reduced gauge-fixed phase space of the system,
which will be denoted by $\Gamma_{0}$
\begin{equation}
\Gamma_{0}:=\left\{  u^{A}\in\Gamma\,|\,\gamma_{I}(u)=0,\,I=1,\ldots
,2n\right\}  \,. \label{Gamma-zero}%
\end{equation}
In $\Gamma_{0}$ a new Poisson structure is introduced by the Dirac bracket
$[\;\;,\;\;]^{\ast}$
\begin{equation}
\lbrack M,N]^{\ast}=[M,N]-[M,\gamma_{I}]C^{IJ}[\gamma_{J},N]\;, \label{db}%
\end{equation}
where $C^{IJ}$ is the inverse of the Dirac matrix constructed from the second
class constraints $\{\gamma_{I}\}$,
\begin{equation}
C_{IJ}=[\gamma_{I},\gamma_{J}]=\Omega^{AB}\partial_{A}\gamma_{I}\partial
_{B}\gamma_{J}\;. \label{d}%
\end{equation}
The symplectic form for the gauge fixed system in the reduced phase space
defines the Dirac bracket (\ref{db}). Suppose now that the set of gauge
conditions $\left\{  G_{i}\right\}  $ fails to fix completely the gauge in a
region of phase space, leading to a Gribov ambiguity (see
Figure\ref{fig_gauge_orbits} ). This means that if a configuration $u^{A}$
satisfies the gauge conditions $G_{i}(u)\approx0$, there exists a
gauge-transformed configuration $u^{A}+\delta u^{A}$ that also satisfies it, namely

%

\begin{equation}
\delta G_{i}\left(  q,p\right)  \approx\partial_{A}G_{i}\delta u^{A}=0\;.
\label{gt}%
\end{equation}
Since gauge transformations are generated by first class constraints,
\begin{equation}
\delta u^{A}=\epsilon^{j}\left[  u^{A},\phi_{j}\right]  \;=\epsilon^{j}%
\Omega^{AB}\partial_{B}\phi_{j}\;, \label{T}%
\end{equation}
where $\epsilon^{j}$ are infinitesimal parameters, the condition for the
existence of Gribov copies (\ref{gt}) takes the form
\begin{equation}
\epsilon^{j}\Omega^{AB}\partial_{A}G_{i}\partial_{B}\phi_{j}=\epsilon
^{j}\left[  G_{i},\phi_{j}\right]  =0\,, \label{FP f}%
\end{equation}
which has nontrivial solutions ($\epsilon^{i}\neq0$) provided
\[
\det\left[  G_{i},\phi_{j}\right]  =0\,.
\]
The matrix $\left[  G_{i},\phi_{j}\right]  $ corresponds to the FP operator in
gauge field theory, whose definition is
\begin{equation}
\mathcal{M}_{ij}=\left[  G_{i},\phi_{j}\right]  =\Omega^{AB}\partial_{A}%
G_{i}\partial_{B}\phi_{j}\,. \label{FP}%
\end{equation}
Gribov ambiguity occurs if the determinant of the FP operator $\mathcal{M}%
_{ij}$ vanishes. The Gribov copies continuously connected to a given
configuration are related by the corresponding zero modes. The GH is defined
to be the subset $\Xi$ of phase space $\Gamma$ where the FP determinant
vanishes,
\begin{equation}
\Xi:=\left\{  u^{A}\in\Gamma\,|\,\det[\mathcal{M}_{ij}]=0\right\}  \,.
\label{g}%
\end{equation}
Now let's observe that the Dirac matrix (\ref{d}) for the set of constraints
$\left\{  \gamma_{I}\right\}  $ contains $\mathcal{M}_{ij}$ as a submatrix
\begin{equation}
C_{IJ}=[\gamma_{I},\gamma_{J}]=\left(
\begin{array}
[c]{cc}%
\Omega^{AB}\partial_{A}G_{i}\partial_{B}G_{j} & \mathcal{M}_{ij}\\
-\mathcal{M}_{ij} & \Omega^{AB}\partial_{A}\phi_{i}\partial_{B}\phi_{j}%
\end{array}
\right)  \,. \label{c}%
\end{equation}
Hence, as the set $\left\{  \phi_{i}\right\}  $ is first class, the
determinant of the Dirac matrix is given weakly by the square of the FP
determinant
\begin{equation}
\det[C_{IJ}]\approx\left(  \det[\mathcal{M}_{ij}]\right)  ^{2}\,. \label{detc}%
\end{equation}
In an open set where $\mathcal{M}_{ij}$ is invertible, the Dirac bracket
(\ref{db}) can safely defined. On the other hand, since at the GH
$\det[\mathcal{M}_{ij}]$ vanishes, the determinant of the Dirac matrix
vanishes as well and the Dirac bracket becomes ill-defined there. Moreover, in
the next section, we will see that a Gribov ambiguity implies a degeneracy of
the symplectic form for the gauge-fixed system at the GH.


\section{Gribov horizon and degenerate surfaces}

In general, the gauge generators $\phi_{i} \approx0$, together with the gauge
fixing conditions $G_{i}\approx0$ form a set of $2n$ second class constraints.
However, this is not globally true in the presence of a Gribov ambiguity,
which can have non-trivial consequences in the symplectic structure of the
reduced phase space. This can be seen considering an open set where the Dirac
matrix $C_{IJ}$ is invertible, $C^{IJ}C_{JK} = \delta^{I}_{K}$. Setting the
constraints strongly to zero defines the reduced gauge-fixed (``physical")
phase space, which is generically a co-dimension $2n$ surface $\Gamma_{0}$
embedded in phase space $\Gamma$.

Even though we started the analysis with a globally invertible symplectic
form, implementing the gauge fixing changes the Poisson structure and a new
symplectic form for the reduced phase space must be found. In order to
explicitly write the symplectic form in the reduced phase space it is useful
to take adapted coordinates $\{U^{A}\}=\{u^{\ast a},v^{I}\}$, where $\{u^{\ast
a}\}$ are \textquotedblleft first class" coordinates (in the sense that they
have vanishing brackets with all second class constraints, see \cite{HT})
\begin{equation}
u^{\ast a}=u^{a}-[u^{a},\gamma_{I}]C^{IJ}\gamma_{J}%
\;\;\mbox{with}\;\;a=1,...,N-2n\;, \label{u2}%
\end{equation}
while $\{v^{I}\}$ is chosen as the set of second class constraints
(\ref{set})
\begin{equation}
v^{I}=\gamma_{I}\;\;\mbox{with}\;\;I=1,...,2n\;. \label{up2}%
\end{equation}
Consequently $\{u^{\ast a}\}$ and $\{v^{I}\}$ are canonically independent
coordinates, i.e.,
\begin{equation}
\lbrack u^{\ast a},v^{I}]=0\,. \label{up3}%
\end{equation}
The conditions $v^{I}=0$ define the reduced phase space, and the $u^{\ast}%
$'s fix the position within the reduced phase space $\Gamma_{0}$. The matrix of their Poisson
brackets given by
\begin{equation}
\hat{\Omega}^{AB}=\left[  U^{A},U^{B}\right]  =\left(
\begin{array}
[c]{cc}%
\omega^{ab} & 0\\
0 & C_{IJ}%
\end{array}
\right)  \,, \label{sp}%
\end{equation}
where
\begin{equation}
\omega^{ab}=[u^{\ast a},u^{\ast b}]\approx\lbrack u^{a},u^{b}]^{\ast}\,,
\label{rs}%
\end{equation}
is the inverse of the symplectic form in the reduced phase space $\omega_{ab}%
$. \newline
The passage from the generic coordinates $\{u^{A}\}$ to the adapted ones,
$\{U^{A}\}=\{u^{\ast a},\gamma_{I}\}$, must be well defined. Then, the
Jacobian for the transformation,
\begin{equation}
\mathcal{J}^{A}{}_{B}=\left(  \frac{\partial U^{A}}{\partial u^{B}}\right)
=\left(
\begin{array}
[c]{c}%
\partial_{B}u^{\ast a}\\
\partial_{B}{\gamma}^{I}%
\end{array}
\right)  , \label{JJ}%
\end{equation}
is invertible. Assuming the original Poisson structure (\ref{cb}) to be well
defined, i.e., $\det[\Omega^{AB}]=\Omega(u)\neq0$, the new Poisson bracket in
the adapted coordinates satisfies
\begin{equation}
\det[\hat{\Omega}^{AB}]=\left(  \det[\mathcal{J}^{A}{}_{B}]\right)  ^{2}%
\Omega\,. \label{detss}%
\end{equation}
Hence, we arrive at the following theorem.\newline

\textbf{Theorem}: \textit{For a system with Gribov ambiguity, the symplectic
form on the reduced phase space, $\omega_{ab}$, necessarily degenerates at the
Gribov horizon.} \newline\textbf{Proof:} Since the coordinates $U^{A}$ are
globally well defined, the determinant of the Jacobian (\ref{JJ}) is finite
everywhere. In particular, it must approach a finite value $\mathcal{J}%
(\bar{u})$ on the GH,
\begin{equation}
\det[\mathcal{J}^{A}{}_{B}]\underset{u\rightarrow\bar{u}}{\longrightarrow
}\mathcal{J}(\bar{u})\neq0\,, \label{det}%
\end{equation}
where $\bar{u}$ stands for the values of the coordinates at the GH (\ref{g}).
From (\ref{sp}) this means that
\begin{equation}
\det[\hat{\Omega}^{AB}]=\det[\omega^{ab}]\det[C_{IJ}]\underset{u\rightarrow
\bar{u}}{\longrightarrow}\mathcal{J}(\bar{u})^{2}\Omega(\bar{u})\,.
\label{detp}%
\end{equation}
On the other hand, from (\ref{detc}) we know that the determinant of the Dirac
matrix vanishes at the GH, and therefore the determinant of the Poisson
structure on the reduced phase space must be singular,
\[
\det[\omega^{ab}]\underset{u\rightarrow\bar{u}}{\longrightarrow}\infty\,.
\]
Consequently, the reduced phase space symplectic form necessarily degenerates
at the GH,
\begin{equation}
\det[\omega_{ab}]\underset{u\rightarrow\bar{u}}{\longrightarrow}0\,.
\label{deg}%
\end{equation}
$\blacksquare$ \newline A well-defined Poisson structure $\omega^{ab}$ at the
GH ($\det[\omega^{ab}(\bar{u})]$ finite) requires $\det[\hat{\Omega}%
^{AB}]\underset{u\rightarrow\bar{u}}{\longrightarrow}0$ and, consequently, the
coordinates $\{U^{A}\}$ should be ill-defined there. This might happen if the
constraints (\ref{set}) are not functionally independent at the GH, that is,
if the constraints fail to be \textit{regular}. If this problem is not
produced by an erroneous choice of gauge fixing, it can only be due to an
irregularity in the first class constraints at the GH. Irregularity in
dynamical systems is an independent problem from degeneracy and requires
special handling to define the system in a consistent manner \cite{MiZ}. An
example of a system with Gribov ambiguity where the reduced symplectic form is
non-degenerate due to irregularities will be analyzed in Section
\ref{irregular}.

The importance of this result is that when the global coordinates are well
defined, the induced symplectic form of the gauge-fixed theory degenerates at
the GH. Consequently, as shown in \cite{STZ} and \cite{dMZ}, the dynamics is
restricted to the regions of phase space bounded by the degeneracy surface.
This argument puts the Gribov-Zwanziger restriction on a firm basis: the
previous analysis (which strictly speaking only holds for finite dimensional
systems) suggests that the system cannot cross the GH (since it is a
degenerate surface for the corresponding Hamiltonian system) and, therefore,
the Gribov-Zwanziger restriction would be naturally respected by the dynamics.


\section{The FLPR Model}

\label{FLPR}

In this section we illustrate the previous discussion with a solvable model
proposed by Friedberg, Lee, Pang and Ren (\textbf{FLPR}), which presents a
Gribov ambiguity for Coulomb-like gauge conditions \cite{flpr}. This model has
been extensively studied trying understand how the Gribov ambiguity could be
circumvented \cite{fujikawa1,banerjee,villanueva}. We will show that, in this gauge, the symplectic form for the gauge-fixed system becomes degenerate at the GH. Closely related models, for which Dirac quantization is non-trivial, have been analized in \cite{Plyushchay}.

\begin{figure}[th]
\begin{center}
\hspace{1cm} \includegraphics[scale=1]{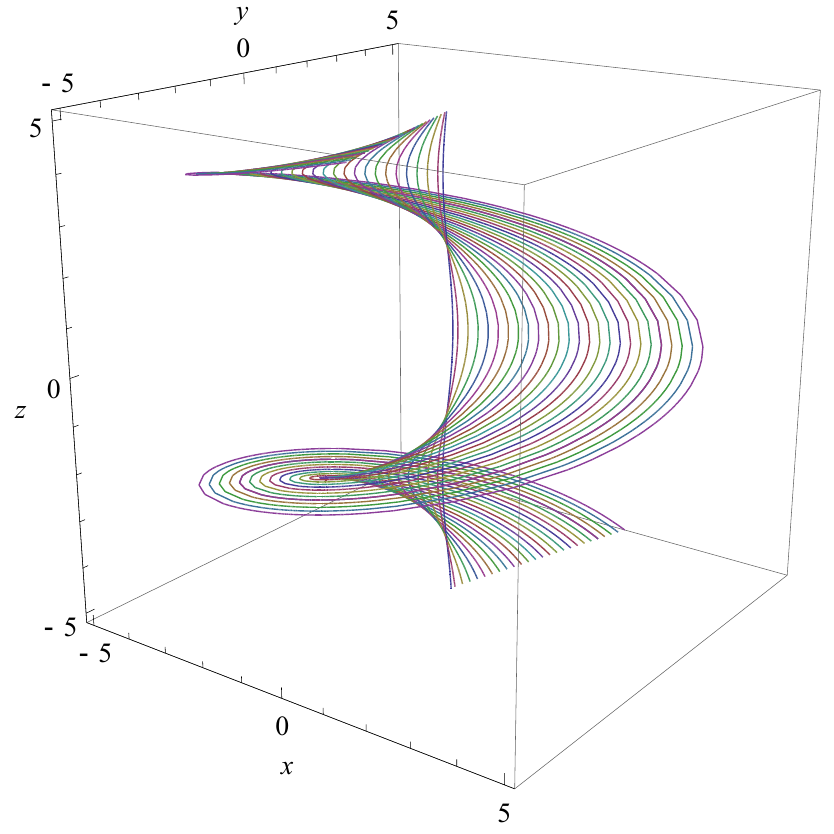}
\end{center}
\caption{The orbits generated by gauge transformations in the FLPR model are
helicoids of the form $(x,y,z)=(\rho\cos[\alpha\varepsilon(t)],\rho\sin
[\alpha\varepsilon(t)],\varepsilon(t))$.}%
\label{fig_helices}%
\end{figure}

The Lagrangian for the FLPR model is
\begin{equation}
L=\frac{1}{2}\left(  (\dot{x}+\alpha y q )^{2} + (\dot{y}-\alpha x q)^{2} +
(\dot{z}-q)^{2} \right)  -V(\rho) \, , \label{L}%
\end{equation}
where $\{x,y,z,q\}$ are Cartesian coordinates, $\rho=\sqrt{x^{2}+y^{2}}$, and
$\alpha>0$ is a coupling constant. The velocity $\dot{q}$ is absent and
therefore the coordinate $q$ plays the role of an auxiliary field or Lagrange
multiplier. The associated canonical momenta are given by
\begin{equation}%
\begin{array}
[c]{ll}%
p_{x} = \frac{\partial L}{\partial\dot{x}} = \dot{x} + \alpha yq\,, & p_{y} =
\frac{\partial L}{\partial\dot{y}} = \dot{y} - \alpha x q\,,\\
p_{z} = \frac{\partial L}{\partial\dot{z}} = \dot{z} - q\,, & p_{q} =
\frac{\partial L}{\partial\dot{q}}=0\,.
\end{array}
\label{p}%
\end{equation}
Following Dirac's procedure, we find one primary constraint
\begin{equation}
\varphi=p_{q}\approx0\,. \label{f}%
\end{equation}
The total Hamiltonian is given by
\begin{equation}
H_{T} = \frac{1}{2}(p_{x}^{2} + p_{y} ^{2} + p_{z} ^{2}) + [\alpha(xp_{y} -
yp_{x}) + p_{z}]q + \xi\varphi+ V(\rho)\,, \label{ht}%
\end{equation}
where $\xi$ is a Lagrange multiplier. Time preservation of the constraint
$\varphi$ leads to the secondary constraint
\begin{equation}
\phi=p_{z}+\alpha\left(  xp_{y}-yp_{x}\right)  \approx0\,, \label{x}%
\end{equation}
which leads to no new constraints for the system. Since $\varphi$ and $\phi$
have vanishing Poisson bracket, they form a first class set, reflecting the
fact that they generate the local\footnote{Locality here refers to time.}
gauge symmetries. The constraint $\varphi$ generates arbitrary translations in
$q$,
\begin{equation}
\delta_{\varphi}(x,y,z,q) = (0,0,0,\varepsilon(t))\,,\;\;\delta_{\varphi
}(p_{x}, p_{y}, p_{z}, p_{q}) = 0\,, \label{varphi}%
\end{equation}
while $\phi$ generates helicoidal motions,
\begin{equation}
\delta_{\phi}(x,y,z,q) = \epsilon(t)(-\alpha y,\alpha x,1,0)\,,\;\;
\delta_{\phi}(p_{x}, p_{y}, p_{z}, p_{q}) = \alpha\epsilon(t)(-p_{y}, p_{x},
0, 0)\,, \label{phi}%
\end{equation}
as it is shown in Figure \ref{fig_helices}. Both transformations leave
invariant the Hamiltonian (\ref{ht}) for arbitrary $\epsilon(t)$ and
$\varepsilon(t)$. Note that the system is invariant under rotations in the
$x-y$ plane, translations in $z$ and time translations, but these are global
symmetries that lead to conservation of the $z$-components of the angular and
the linear momenta, and the energy. Symmetries (\ref{varphi}, \ref{phi}),
instead, are not rigid but local.

The gauge freedom generated by $\varphi$, can be fixed by the gauge condition
\begin{equation}
\mathcal{G}=q\approx0\;,\label{tem}%
\end{equation}
which is analogous to the temporal gauge $A_{0}=0$ in Maxwell theory. Thus,
the coordinate $q$ and its conjugate momentum $p_{q}$, can be eliminated from
phase space by an algebraic gauge choice, as it happens with $A_{0}$ in
electrodynamics, which also enters as a Lagrange multiplier. This partial
gauge fixing eliminates the term $\xi\varphi$ from Hamiltonian (\ref{ht}) and
identifies $q$ as a Lagrange multiplier. The result is a Hamiltonian system in
the 6-dimensional phase space $\Gamma$ with coordinates $\{u^{A}\}=\{x,p_{x}%
,y,p_{y},z,p_{z}\}$ and a single (necessarily first class) constraint
$\phi\approx0$. The Poisson bracket in this phase space is given by
\begin{equation}
\lbrack M,N]_{\Gamma}=\Omega^{AB}\partial_{A}M\partial_{A}N\;,\label{PoissonB}%
\end{equation}
where $\Omega^{AB}$ is the canonical Poisson bracket, and the canonical
symplectic form is
\begin{equation}
\Omega_{AB}=\left(
\begin{array}
[c]{rrrrrr}%
0 & 1 & 0 & 0 & 0 & 0\\
-1 & 0 & 0 & 0 & 0 & 0\\
0 & 0 & 0 & 1 & 0 & 0\\
0 & 0 & -1 & 0 & 0 & 0\\
0 & 0 & 0 & 0 & 0 & 1\\
0 & 0 & 0 & 0 & -1 & 0
\end{array}
\right)  \,.
\end{equation}

\begin{figure}[th]
\begin{center}
\hspace{1.2cm} \includegraphics[scale=0.85]{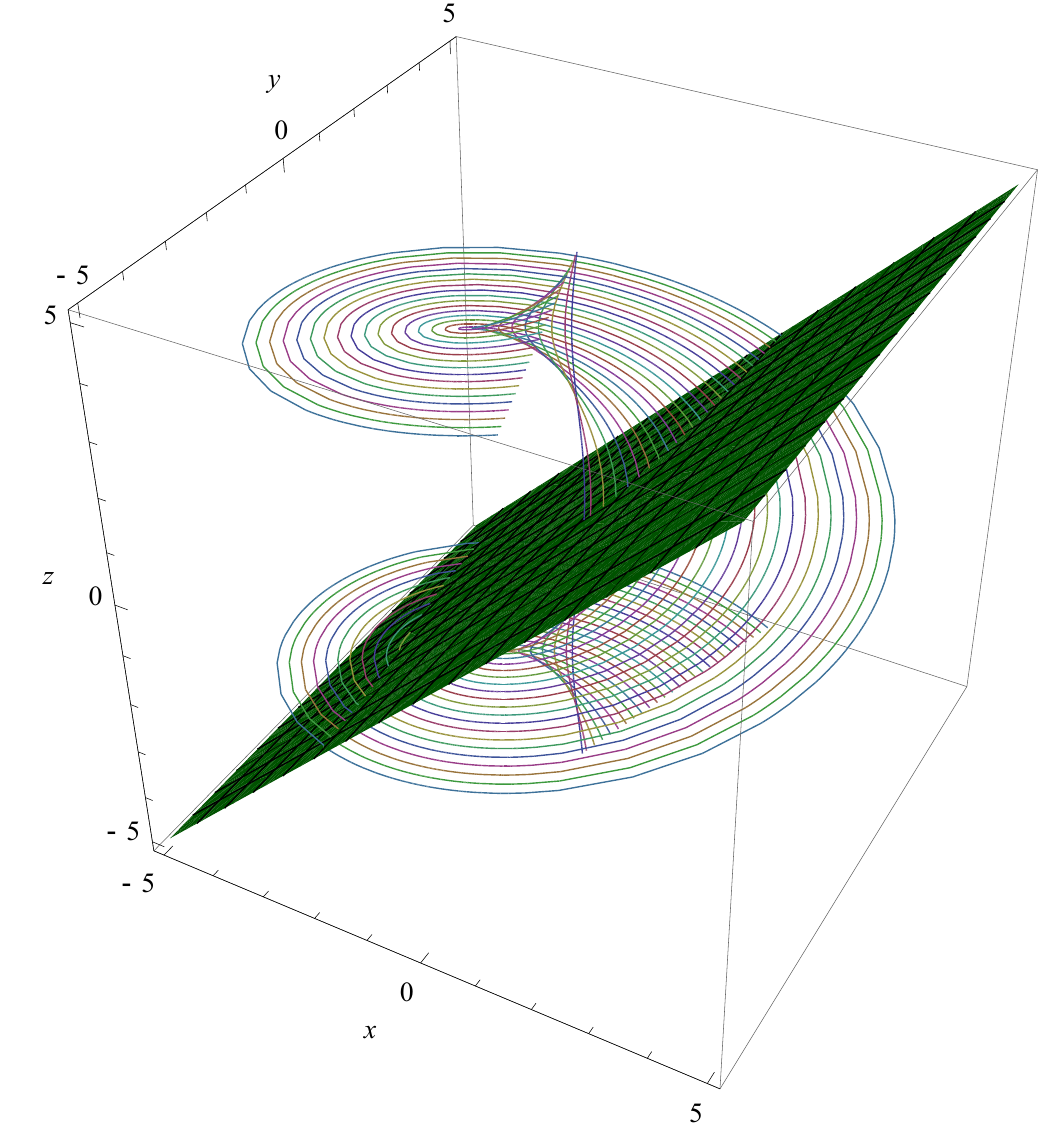}
\end{center}
\caption{ The surface defined by the $\lambda$-gauge condition $G=z-\lambda
x=0$ is a plane (here plotted for $\lambda=1$). The Gribov ambiguity in the
FLPR model is reflected by the fact that this plane intersects some gauge
orbits more than once.}%
\label{fig_gauge_fixed_plan}%
\end{figure}

Following \cite{flpr}, the gauge freedom generated by $\phi$ is to be
eliminated by a gauge condition $G(x,y,z)\approx0$, where $G$ is a linear
homogeneous function, which is in some sense analogous to the Coulomb gauge.
Since the system is invariant under rotations in the $x$-$y$ plane, we can
choose the gauge condition to be independent of $y$. Hence, we take
\begin{equation}
G=z-\lambda x\approx0\;, \label{coulomb}%
\end{equation}
which is called \textquotedblleft$\lambda$-gauge". As it can be seen, for
$\lambda\neq0$ the condition (\ref{coulomb}) does not fix the gauge globally
(see Figure \ref{fig_gauge_fixed_plan}). In the same way as the Coulomb gauge
does in Yang-Mills theory, it has a Gribov ambiguity at $y=-(\alpha
\lambda)^{-1}$. In fact, the non trivial Poisson bracket,
\begin{equation}
\mathcal{M}=[G,\phi]=1+\alpha\lambda y\,, \label{M=gphi}%
\end{equation}
which corresponds to the Faddeev-Popov determinant, indicates that these are
second class constraints everywhere in $\Gamma_{0}$, except at $y=-(\alpha
\lambda)^{-1}$. Consequently, $\det C=\mathcal{M}^{2}$ vanishes where the
condition $G\approx0$ fails to fix the gauge, that is on the Gribov horizon
\begin{equation}
\Xi=\{(x,p_{x},y,p_{y},z,p_{z})\in\Gamma\,|\mathcal{M}=0\}\,.
\label{g-horizon}%
\end{equation}

When the second class constraints (\ref{x}, \ref{coulomb}) are set strongly
equal to zero, $z$ and $p_{z}$ can be eliminated from the phase space. The
four-dimensional reduced phase space $\Gamma_{0}$, parametrized with
coordinates $(x,p_{x},y,p_{y})$, acquires a non-canonical Poisson structure
given by the Dirac bracket (\ref{db}) , where $\gamma_{I}$ are the second
class constraints $\{G,\phi\}$
\begin{equation}
\gamma_{I}:\;\gamma_{1}=G=z-\lambda x\;,\qquad\gamma_{2}=\phi=p_{z}%
+\alpha(xp_{y}-yp_{x})\;,
\end{equation}
and $C^{IJ}$ is the inverse of the Dirac matrix $C_{IJ}\equiv\lbrack\gamma
_{I},\gamma_{J}]$ . In this case, the Dirac matrix is given by
\begin{equation}
C_{IJ}=[\gamma_{I},\gamma_{J}]_{\Gamma}=\left(
\begin{array}
[c]{cc}%
0 & \mathcal{M}\\
-\mathcal{M} & 0
\end{array}
\right)  \;, \label{dm}%
\end{equation}
and the Dirac brackets are given by
\begin{align}
\lbrack x,p_{x}]^{\ast}  &  =\frac{1}{\mathcal{M}}\;\;,\;\;[x,y]^{\ast
}=0\;\;,\;\;[x,p_{y}]^{\ast}=0\;,\label{D-brackets}\\
\lbrack y,p_{y}]^{\ast}  &  =1\;\;,\;\;[y,p_{x}]^{\ast}=\frac{\alpha\lambda
x}{\mathcal{M}}\;\;,\;\;[p_{x},p_{y}]^{\ast}=-\frac{\alpha\lambda p_{x}%
}{\mathcal{M}}\;.\nonumber
\end{align}

In the reduced phase space, the Poisson matrix (\ref{rs}) takes the form
\begin{equation}
\omega^{ab}=\left(
\begin{tabular}
[c]{cccc}%
$0$ & $\frac{1}{\mathcal{M}}$ & $0$ & $0$\\
$-\frac{1}{\mathcal{M}}$ & $0$ & $-\frac{\alpha\lambda x}{\mathcal{M}}$ &
$-\frac{\alpha\lambda p_{x}}{\mathcal{M}}$\\
$0$ & $\frac{\alpha\lambda x}{\mathcal{M}}$ & $0$ & $1$\\
$0$ & $\frac{\alpha\lambda p_{x}}{\mathcal{M}}$ & $-1$ & $0$%
\end{tabular}
\ \ \ \ \ \ \ \right)  \,, \label{pf}%
\end{equation}
and the corresponding symplectic form is
\begin{equation}
\omega_{ab}=\left(
\begin{array}
[c]{cccc}%
0 & -\mathcal{M} & -\alpha\lambda p_{x} & \alpha\lambda x\\
\mathcal{M} & 0 & 0 & 0\\
\alpha\lambda p_{x} & 0 & 0 & -1\\
-\alpha\lambda x & 0 & 1 & 0
\end{array}
\right)  \,. \label{sympf}%
\end{equation}
It can be checked that this symplectic form is closed, $\partial_{a}%
\omega_{bc}+\partial_{b}\omega_{ca}+\partial_{c}\omega_{ab}=0$ and therefore
in a local chart it can be expressed as the exterior derivative of a one-form,
$\omega_{ab}=\partial_{a}X_{b}-\partial_{b}X_{a}$ (or $\omega=dX$), which can
be integrated as
\begin{equation}
X(x,p_{x},y,p_{y})=(p_{x}+\alpha\lambda\lbrack yp_{x}-xp_{y}])dx+p_{y}dy\,.
\label{X}%
\end{equation}
The determinant of the symplectic form in the reduced phase space can be read
off from (\ref{sympf}), and is given by%
\begin{equation}
\det[\omega_{ab}]=\mathcal{M}^{2}\,. \label{detredsimp}%
\end{equation}
Clearly, $\omega_{ab}$ degenerates precisely at the Gribov (\ref{g-horizon})
restricted to the constraint surface and the degeneracy surface
(\ref{degeneracy}) is given by
\begin{equation}
\Sigma=\{(x,p_{x},y,p_{y})\in\Gamma_{0}|\,\Upsilon(u)\equiv\mathcal{M}=0\}\,.
\label{degflpr}%
\end{equation}
This corresponds to a particular realization of the behavior (\ref{deg}). In
fact, defining $\sigma^{2}=1+\alpha^{2}\lambda^{2}\rho^{2}>0$, the eigenvalues
of the reduced symplectic form are given by $\{\pm i\omega_{+},\pm i\omega
_{-}\}$, where
\begin{equation}
\omega_{\pm}=\frac{1}{\sqrt{2}}\left[  \sigma^{2}+\mathcal{M}^{2}\pm
\sqrt{(\sigma^{2}+\mathcal{M}^{2})^{2}-4\mathcal{M}^{2}}\right]  ^{1/2}.
\end{equation}
Near the degeneracy $\omega_{+}$ and $\omega_{-}$ can be expanded in powers of
$\mathcal{M}$, leading to
\begin{equation}
\omega_{+}\approx\sigma\text{ \ \ }\;\;,\;\;{\ \ }\omega_{-}\approx
\frac{\mathcal{M}}{\sigma}\;. \label{eigen}%
\end{equation}
Hence, as the system approaches to the degeneracy $\omega_{+}$ goes linearly
to zero while $\omega_{-}$ never vanishes, which means that the symplectic
form $\omega_{ab}$ has a simple zero in the degeneracy surface and this system
corresponds to the class of degenerate systems discussed in \cite{STZ} and
\cite{dMZ}.

It is reassuring to confirm that the degeneracy is not an artifact introduced
by the change of coordinates $\{U^{A}\}\rightarrow\{u^{\ast a},v^{I}\}$
defined in (\ref{u2}), which in this case is given by
\begin{equation}%
\begin{array}
[c]{ll}%
x^{\ast}=\dfrac{x+\alpha yz}{\mathcal{M}}\,, & p_{x}^{\ast}=\dfrac
{p_{x}+\alpha p_{y}z+\alpha\lambda p_{z}}{\mathcal{M}}\,,\\
y^{\ast}=y-\dfrac{\alpha x(z-\lambda x)}{\mathcal{M}}\,, & p_{y}^{\ast}%
=p_{y}-\dfrac{\alpha p_{x}(z-\lambda x)}{\mathcal{M}}\,,\\
v^{1}=\gamma_{1}=z-\lambda x\,, & v^{2}=\gamma_{2}p_{z}+\alpha\left(
xp_{y}-yp_{x}\right)  \,.
\end{array}
\,, \label{u*a}%
\end{equation}
In fact, the Jacobian (\ref{JJ}) is given in this case by
\begin{equation}
\mathcal{J}^{A}{}_{B}=\left(
\begin{array}
[c]{cccccc}%
\frac{1}{\mathcal{M}} & 0 & 0 & 0 & \frac{\alpha y}{\mathcal{M}} & 0\\
0 & \frac{1}{\mathcal{M}} & -\frac{\alpha\lambda p_{x}}{\mathcal{M}} &
\frac{\alpha\lambda x}{\mathcal{M}} & \frac{\alpha\lambda p_{y}}{\mathcal{M}}
& \frac{\lambda}{\mathcal{M}}\\
\frac{\alpha\lambda x}{\mathcal{M}} & 0 & 1 & 0 & -\frac{\alpha x}%
{\mathcal{M}} & 0\\
\frac{\alpha\lambda p_{x}}{\mathcal{M}} & 0 & 0 & 1 & -\frac{\alpha\lambda
p_{x}}{\mathcal{M}} & 0\\
-\lambda & 0 & 0 & 0 & 1 & 0\\
\alpha p_{y} & -\alpha y & -\alpha p_{x} & \alpha x & 0 & 1
\end{array}
\right)  \,, \label{jf}%
\end{equation}
which, in spite of the the apparent singularities in its entries, has unit
determinant everywhere in phase space, $(\det\mathcal{J})|_{\Gamma}\equiv1$.


\subsection{Effective Lagrangian for the gauge-fixed system}

The gauge-fixed system is a degenerate one described by a first order
Hamiltonian action, as presented in (\ref{first_order_action}),
\begin{equation}
I_{gf}[u]=\int\mathrm{d}t\,[\dot{u}^{a}X_{a}(u)-H_{gf}(u)]\, ,
\end{equation}
where $X_{a}$ is given by (\ref{X}), $H_{gf}$ is the gauge-fixed Hamiltonian,
\begin{align}
H_{gf}  &  =\frac{1}{2}(1 + \alpha^{2} y^{2}) p_{x}^{2} + \frac{1}{2}(1 +
\alpha^{2} x^{2}) p_{y}^{2} - \alpha^{2} xy p_{x} p_{y} + V(x^{2} +
y^{2})\nonumber\\
&  =\frac{1}{2} g^{ij}p_{i} p_{j} + V(x^{2} + y^{2})\, . \label{hgff}%
\end{align}
Here the matrix
\begin{equation}
g^{ij}:= \left[
\begin{array}
[c]{cc}%
(1+\alpha^{2} y^{2}) & -\alpha^{2} xy\\
-\alpha^{2} xy & (1+\alpha^{2} x^{2})
\end{array}
\right]  \, \label{inv-metric}%
\end{equation}
is the inverse of the metric
\begin{equation}
g_{ij}:= \frac{1}{1+\alpha^{2} \rho^{2}}\left[
\begin{array}
[c]{cc}%
(1+\alpha^{2} x^{2} ) & \alpha^{2} xy\\
\alpha^{2} xy & (1+\alpha^{2} y^{2})
\end{array}
\right]  . \label{metric}%
\end{equation}


\subsection{Gauge Orbits and Phase-Space}

Gribov ambiguity results from the fact that the surface defined by a gauge
condition does not intersect every gauge orbit once and only once. As it was
mentioned in Section \ref{gf}, this is a requirement to achieve a proper gauge
fixing \cite{HT}. In the case of the FLPR model this clearly happens because
the plane defined by (\ref{coulomb}) intersects some gauge orbits many times
for $\lambda>0$, as it can be seen in Figure \ref{fig_gauge_fixed_plan}. The
$G=0$ plane intersects more than once any orbit such that $x^{2}+y^{2}%
>(\alpha\lambda)^{-2}$. The only way that this doesn't happen is if
$\lambda=0$.

\begin{figure}[th]
\begin{center}
\hspace{1cm} \includegraphics[scale=0.85]{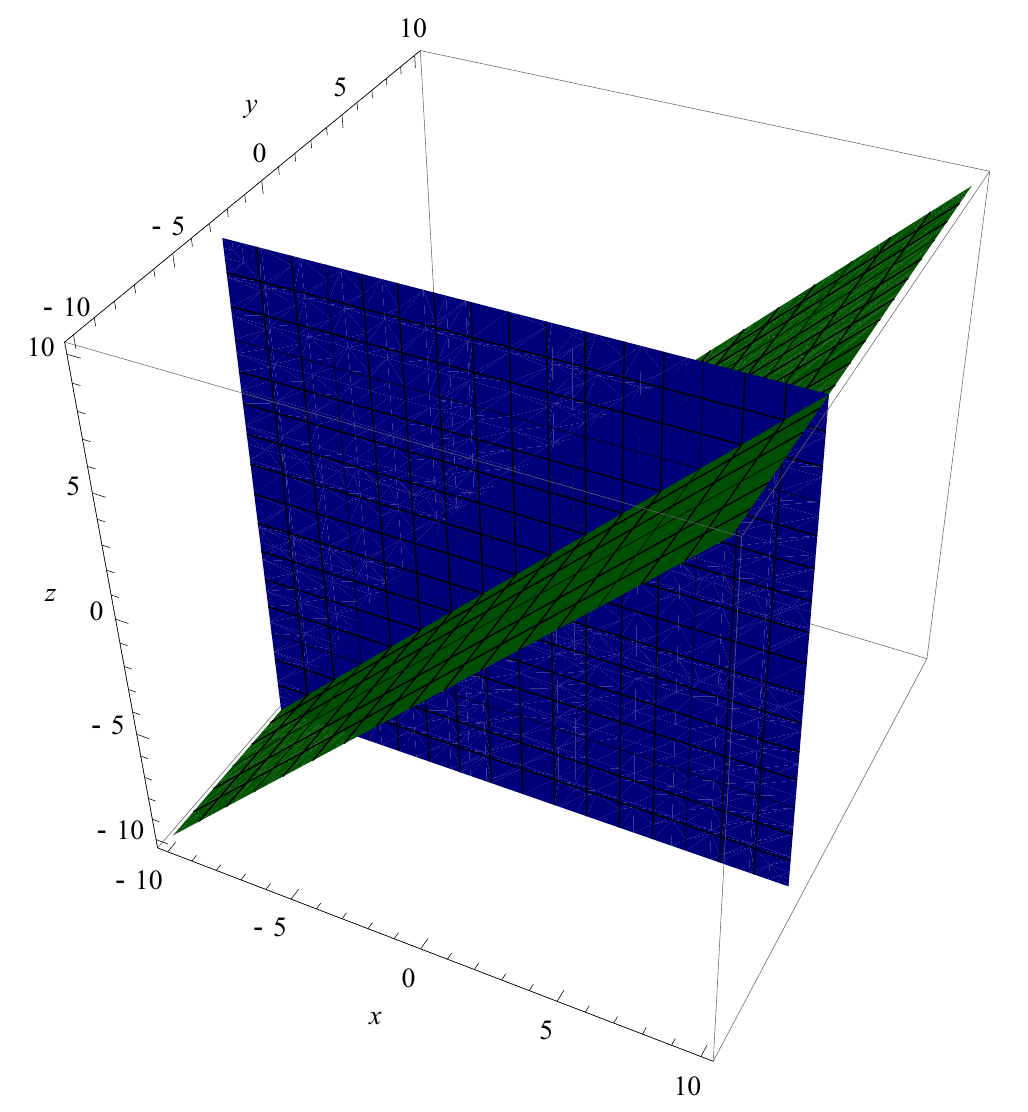}
\end{center}
\caption{In the case of the FLPR model, the Gribov horizon (blue plane),
$y=-\left(  \alpha\lambda\right)  ^{-1}$, and the constraint surface (green
plane), $G=0$, are plotted for $\lambda=1$ and $\alpha=1/3$. The GH divides
the constraint surface in two dynamically disconnected regions.}%
\label{fig_degeneracy_plan}%
\end{figure}

Degenerate surfaces divide phase space into dynamically disconnected regions.
In this case the presence of the GH defines two regions in physical gauge
fixed space (see Figure \ref{fig_degeneracy_plan})
\begin{align}
C_{+}  &  :=\left\{  \left(  x,y, z\right)  | \, z- \lambda x=0, \,
1+\alpha\lambda y>0\right\}  \,,\\
C_{-}  &  :=\left\{  \left(  x,y,z\right)  |\, z- \lambda x=0, \,
1+\alpha\lambda y<0\right\}  \,.
\end{align}
These two regions are not equivalent since only $C_{+}$ contains at least one
representative of every gauge orbit, while not all gauge orbits pass through
$C_{-}$. To restrict the analysis of the system to one region or the other is
consistent in the sense that all states whose initial condition is in one
region will remain there always (see \cite{dMZ}). In Yang-Mills theories the
Gribov region corresponds to the neighborhood of $A_{\mu}=0$ in the functional
space of connections where the FP operator is positive definite \cite{DeW03}
and \textquotedblleft small copies\textquotedblright\ (namely, points
infinitesimally close which belong to the same gauge orbit) are absent. In the
Yang-Mills case all the gauge orbits cross the Gribov region at least once
\cite{DZ89}. Similarly to what happens in the Yang-Mills case, within the
region $C_{+}$ (which contains at least one representative of every gauge
orbit) there are still large copies\footnote{We remind that large copies are
points belonging to the same gauge orbit (and, of course, satisfying the same
gauge condition) which are not infinitesimally close. This means that large
copies (unlike the small ones) do not correspond to zero-modes of the
Faddeev-Popov operator. In Yang-Mills theory, the pattern of appearance of the
large Gribov copies within the Gribov region is very complicated and only few
examples are known \cite{Va92}.}.


\subsection{Quantization}

In order to define the quantum theory, the Hilbert space for the system\ must
be equipped with an inner product that provides a scalar product and a norm
\begin{equation}
\left\Vert \psi(u) \right\Vert =\left(  \int\mathrm{d}^{2} u \, \sqrt{g}\,w(u)
\left\vert \psi(u) \right\vert ^{2} \right)  ^{1/2}. \label{norm2}%
\end{equation}

In the FLPR model, $g=(1 + \alpha^{2} \rho^{2} )^{-1}$ is the determinant of
the metric (\ref{metric}) and the weight $w(u)$ is such that the Hamiltonian
is symmetric, that is,
\begin{equation}
\int\mathrm{d}^{2}u \, \sqrt{g} \, w(u)\psi_{1}^{*}(u) \left(  \hat{H}\psi
_{2}(u) \right)  =\int\mathrm{d}^{2} u \, \sqrt{g} \, w(u)\left(  \hat{H}%
\psi_{1} (u) \right)  ^{*}\psi_{2} (u)\, .
\end{equation}
As discussed in Section \ref{sec_deg_syst}, the proper choice for the measure
$w(u)$ corresponds to the Pfaffian of the symplectic form $\omega_{ab}$
(\ref{sympf}), given in this case by (\ref{degflpr}),
\begin{equation}
w(u) = \Upsilon=\mathcal{M} = 1+\alpha\lambda y\; . \label{w}%
\end{equation}
whose zeros define the degeneracy surface (\ref{degeneracy}). In order to see this, let's define new variables $\{\pi_{i} \}$ canonically
conjugate to the $u$'s, so that
\begin{equation}
[u^{i} ,\pi_{j} ]^{*} = \delta_{j}^{i}, \quad\left\{  u^{i} \right\}  =
\left\{  x,y\right\}  \, . \label{pi}%
\end{equation}
A simple calculation using (\ref{D-brackets}) leads to the following
expression of the momenta in terms of the $\pi$'s
\begin{equation}
p_{x} =\frac{1}{1+\alpha\lambda y} (\pi_{x}+ \alpha\lambda x \pi_{y})\,, \quad
p_{y} =\pi_{y}\, . \label{pp}%
\end{equation}
The quantum operators are then obtained via the prescription
\begin{align}
u^{i}  &  \longrightarrow\hat{u}^{i} = u^{i}\,,\nonumber\\
\pi_{i}  &  \longrightarrow\hat{\pi}_{i} = -i\hbar\partial_{i}\,,\label{qp}\\
[ \text{ },\text{ }]^{*}  &  \longrightarrow\frac{1}{i\hbar} \left[  \text{ },
\text{ } \right]  \mbox{(Commutator)}\,.\nonumber
\end{align}
Using (\ref{pp}), the classical Hamiltonian (\ref{hgff}) can be rewritten as
\begin{equation}
H =\frac{1}{2}h^{ij}\pi_{i}\pi_{j}+V \, , \label{h1}%
\end{equation}
where $h^{ij}$ is the inverse of the metric
\begin{equation}
h_{ij}=\frac{1}{1+\alpha^{2} \rho^{2} }\left(
\begin{array}
[c]{cc}%
\left(  1+\alpha\lambda y\right)  ^{2} + \alpha^{2} \left(  1+ \lambda^{2}
\right)  x^{2} & \alpha^{2} xy - \alpha\lambda x\\
\alpha^{2} xy - \alpha\lambda x & 1+\alpha^{2} y^{2}%
\end{array}
\right)  . \label{met}%
\end{equation}

At the quantum level, the correct ordering for the quantum operators
(\ref{qp}) that renders the Hamiltonian symmetric --and invariant under
general coordinate transformations-- is the one for which $\hat{H}$ is a
Laplacian for the metric $h_{ij}$ \cite{ChrisZ}, i.e.
\begin{equation}
\hat{H}=-\frac{\hbar^{2}}{2}\frac{1}{\sqrt{\left\vert h\right\vert }}%
\partial_{i}\left(  \sqrt{\left\vert h\right\vert }h^{ij}\partial_{j}\right)
+V\left(  \rho\right)  \,,
\end{equation}
where $h$ is the determinant of (\ref{met}) and where integration measure in
(\ref{norm2}) is $\int d^{2}u\sqrt{h}$. A straightforward computation leads
to
\begin{equation}
\sqrt{h}=\frac{(1+\alpha\lambda y)}{\sqrt{1+\alpha^{2}\rho^{2}}}=\sqrt
{g}\Upsilon\,,
\end{equation}
which confirms (\ref{w}). Hence, the measure of the Hilbert space vanishes
exactly where the symplectic form does. Then, according to the results in
\cite{dMZ} this permits to interpret the corresponding Hilbert space as a
collection of causally disconnected subspaces: there is no tunneling from one
side of the degenerate surface to the other. In turn, this confirms the
dynamical correctness of imposing the restriction to the interior of the
Gribov region, at least for first quantization.


\section{Irregular case}

\label{irregular}

As mentioned in Section 4, there is an exceptional case in which the reduced
symplectic form is non-degenerate at the GH. As it will be shown in the
following, this could happen if the constraints fail to be functionally
independent, i.e., if they are irregular \cite{HT, MiZ}.

A set of constraints is regular if they are functionally independent on the
constraints surface. For a set of constraints (\ref{set}) this is ensured by
demanding that the Jacobian
\begin{equation}
\mathcal{K}^{I}{}_{B}=\left.  \frac{\partial\gamma_{I}}{\partial u^{B}%
}\right\vert _{\Gamma_{0}}\label{k}%
\end{equation}
has maximal rank. In particular, for a set of two constraints $\{G,\phi\}$,
this means
\begin{equation}
dG\wedge d\phi|_{\Gamma_{0}}\neq0\Longrightarrow\partial_{\lbrack A}%
G\partial_{B]}\phi|_{\Gamma_{0}}\neq0\,,\label{reg3}%
\end{equation}
while the Dirac matrix (\ref{c}) takes the form%
\begin{equation}
C_{IJ}=\left(
\begin{array}
[c]{cc}%
0 & \mathcal{M}\\
-\mathcal{M} & 0
\end{array}
\right)  \,,\label{diraccl}%
\end{equation}
where $\mathcal{M}=\left[  G,\phi\right]  $ the FP determinant. Hence, using
(\ref{db}) and (\ref{c}), the reduced phase space symplectic form (\ref{rs})
can be expressed weakly as
\begin{equation}
\omega^{ab}\approx\lbrack u^{a},u^{b}]^{\ast}=\Omega^{ab}+\mathcal{M}%
^{-1}\Omega^{aC}\Omega^{Db}\partial_{\lbrack C}G\partial_{D]}\phi\,.
\end{equation}
This suggests that, if the constraints fail the regularity test (\ref{reg3})
at the GH, the singularity in the inverse of the FP determinant $\mathcal{M}%
^{-1}$ can be cancelled by the vanishing quantity $\partial_{\lbrack
C}G\partial_{D]}\phi$ and no degeneracies would appear even in the presence of
Gribov ambiguity.

Another way to see this picture for a general set of constraints (\ref{set}),
$\left\{  \gamma_{I}\right\}  =$ $\{G_{i},\phi_{j}\}$, is by noting that, as
the original symplectic strucutre (\ref{cb}) is considered to be well defined
($\det[\Omega^{AB}]=\Omega$), the determinant of the Poisson bracket in the
new coordinates $U^{A}=[u^{\ast a},v^{I}]$, defined by (\ref{u2}) and
(\ref{up2}) is given by (\ref{detss}), which can be evaluated on the
constraint surface $\Gamma_{0}$,
\begin{equation}
\left.  \det[\hat{\Omega}^{AB}]\right\vert _{\Gamma_{0}}=\left.  \left(
\det\left[  \mathcal{J}^{A}{}_{B}\right]  \right)  ^{2}\Omega\right\vert
_{\Gamma_{0}}.
\end{equation}
On the other hand, the Jacobian (\ref{JJ}) evaluated on $\gamma_{I}=0$ can be
written in terms of (\ref{k}) as
\begin{equation}
\left.  \mathcal{J}^{A}{}_{B}\right\vert _{\Gamma_{0}}=\left(
\begin{array}
[c]{c}%
\partial_{B}u^{\ast a}\\
\mathcal{K}^{I}{}_{B}%
\end{array}
\right)  \,. \label{jacobian_gradient}%
\end{equation}
Hence, if the constraints (\ref{set}) are irregular at the GH, both
$\mathcal{K}^{I}{}_{B}$ and $\mathcal{J}^{A}{}_{B}|_{\Gamma}$ have non-maximal
rank, implying that the determinant $\det\left[  \mathcal{J}^{A}{}_{B}\right]
$ vanishes at the intersection of the GH and $\Gamma_{0}$. Therefore,
\begin{equation}
\left.  \det[\hat{\Omega}^{AB}]\right\vert _{\Gamma_{0}}\underset
{u\rightarrow\bar{u}}{\longrightarrow}0\,. \label{detp2}%
\end{equation}
Then, looking again at (\ref{sp}), we see that in this case the vanishing of
$\det\left[  C_{IJ}\right]  $ at the GH does not imply that the reduced phase
space Poisson structure should blow up and degeneracies in the symplectic
structure of the gauge fixed system can be overcome. However, this situation
is even more pathological than the degenerate one, as the gauge-fixed system
doesn't describe the dynamics of the original system. In the following an
explicit example of this situation will be presented.


\subsection{Example: Christ-Lee Model}

The Lagrangian for the Christ-Lee model \cite{CL} is given by
\[
L=\frac{1}{2}(\dot{x}+\alpha yq)^{2}+(\dot{y}-\alpha xq)^{2}-V(x^{2}%
+y^{2})\,,
\]
where $\alpha>0$ is a coupling constant. The canonical momenta of the system
are given by
\begin{equation}%
\begin{array}
[c]{lll}%
p_{x}=\dot{x}+\alpha yq\,, & p_{y}=\dot{y}-\alpha xq\,, & p_{q}=0\,.
\end{array}
\end{equation}
Dirac's method leads to the following first class constraints
\begin{equation}
\varphi=p_{q}\approx0\,,\qquad\phi=xp_{y}-yp_{x}\approx0\,,
\end{equation}
which generate arbitrary translations in $q$ and rotations in the $x-y$ plane
respectively. The total Hamiltonian is given by
\begin{equation}
H_{T}=\frac{1}{2}(p_{x}^{2}+p_{y}^{2})+\alpha(xp_{y}-yp_{x})q+\xi
\varphi+V(\rho)\,, \label{ht_christ}%
\end{equation}
where $\xi$ is a Lagrange multiplier As before, the constraint $\varphi$ can
be trivially eliminated by the introduction of a gauge condition
$\mathcal{G}=q\approx0$. The Dirac bracket associated to this pair of
constraints is just the Poisson bracket in the coordinates $\{x,p_{x}%
,y,p_{y}\}$, and using this we can set $\varphi$ and $\mathcal{G}$ strongly to
zero. Now we will focus on the constraint $\phi$, whose action on the
coordinates generates circular orbits in phase space.

\begin{figure}[th]
\begin{center}
\hspace{1cm} \includegraphics[scale=0.85]{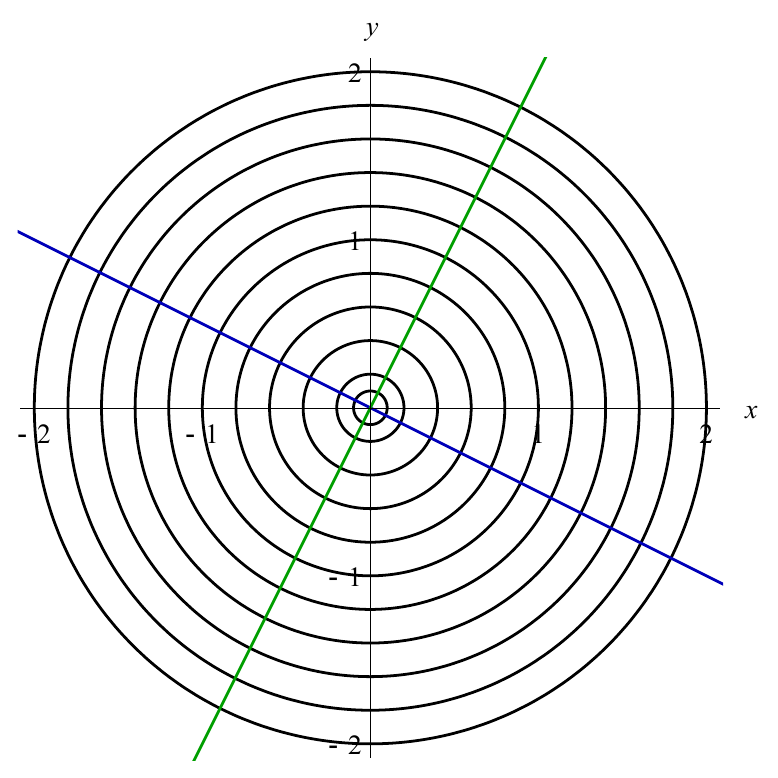}
\end{center}
\caption{ Orbits for the Christ-Lee model are given by circles centered at the
origin. The GH (blue line) and the surface $G=0$ (green line) are plotted for
$\mu=2$. The GH restricted to the constraint surface corresponds to the point
$x=y=0$.}%
\label{fig_degeneracy}%
\end{figure}

As we are interested in Gribov ambiguity, we will pick the following gauge
condition \cite{banerjee}
\begin{equation}
G=y-\mu x\approx0\,,
\end{equation}
with $\mu$ a constant. The Dirac matrix for this set of constraints
$\gamma_{I}=\left\{  G,\phi\right\}  $ with $I=1,2$ is given by (\ref{diraccl}%
) where%
\[
\mathcal{M}=\left[  G,\phi\right]  =x+\mu y
\]
and there exists a GH (\ref{g}) defined by
\begin{equation}
\Xi:=\left\{  \left(  x,p_{x},y,p_{y}\right)  \in\Gamma\,|\,\mathcal{M}%
=0\right\}  \,.\label{gcl}%
\end{equation}

The Poisson structure of the space is given via the Dirac bracket (\ref{db}),
where $\gamma_{I}$ are the second class constraints $\{G,\phi\}$. This leds to%
\begin{equation}%
\begin{array}
[c]{c}%
\lbrack x,p_{x}]^{\ast}=\dfrac{x}{\mathcal{M}}\;\;,\;\;[x,y]^{\ast
}=0\;\;,\;\;[x,p_{y}]^{\ast}=\dfrac{y}{\mathcal{M}}\,,\\
\lbrack y,p_{y}]^{\ast}=\dfrac{\mu y}{\mathcal{M}}\;\;,\;\;[y,p_{x}]^{\ast
}=\dfrac{-\mu y}{\mathcal{M}}\;\;,\;\;[p_{x},p_{y}]^{\ast}=\dfrac{\mu
p_{x}p_{y}}{\mathcal{M}^{2}}\,.
\end{array}
\label{dirac_lee}%
\end{equation}
Once the second class constraints $G$ and $\phi$ are strongly equal to zero
\begin{equation}
y=\mu x\;\;,\;\;\quad\quad p_{y}=\mu p_{x}\;, \label{cfl}%
\end{equation}
we are left with only one degree of freedom corresponding to the variable $x$.
The Gribov horizon restricted to the constraint surface $G=0$ is given by
$x=0$ (see Figure \ref{fig_degeneracy}). Then, the reduced phase space symplectic form (\ref{rs}) turns out to be
non-degenerate
\begin{equation}
\omega_{ab}=\left(
\begin{array}
[c]{cc}%
0 & -\left(  1+\mu^{2}\right) \\
1+\mu^{2} & 0
\end{array}
\right)  \;\;,\;\;\det[\omega_{ab}]=\left(  1+\mu^{2}\right)  ^{2}\;.
\end{equation}
However, in this case, the constraints $\{G,\phi\}$ are not functionally
independent at the GH. To see this consider the sub-block (\ref{k}) of
(\ref{jacobian_gradient}) whose rank determines the functional independence of
the constraints $\{G,\phi\}$,
\begin{equation}
\mathcal{K}^{I}{}_{B}=\left.  \frac{\partial\gamma_{I}}{\partial u^{B}%
}\right\vert _{\Gamma_{0}}=\left(
\begin{array}
[c]{cccc}%
-\mu & 0 & 1 & 0\\
\mu p_{x} & -\mu x & -p_{x} & x
\end{array}
\right)  \;.
\end{equation}
This matrix has non-maximal rank on the GH restricted to the constraint
surface ($x=0$), then the constraints are not regular there because their
gradients are proportional. 

The gauge-fixed Lagrangian now reads
\begin{equation}
L=\frac{1}{2}\left(  1+\mu^{2}\right)  \dot{x}^{2}-V\left(  (1+\mu^{2}%
)x^{2}\right)  \,,
\end{equation}
which seems to be free of degeneracy at the GH. However this is an illusion
because the absence of degeneracy results from the fact that the constraints
are no longer functionally independent, so that the system, on the Gribov
horizon, fails to be regular.

\section{Conclusions and further comments}

We have discussed the relation between Gribov ambiguity and degeneracy in
Hamiltonian systems. In our analysis, the Gribov-Zwanziger restriction can be
seen as a prescription consistent with the fact that it is respected by the
dynamics, both classical and quantum mechanically, at least in finite
dimensional Hamiltonian systems.

In gauge systems with finite number of degrees of freedom, the existence of
Gribov ambiguity in the gauge fixing conditions leads to a degenerate
symplectic structure for the reduced system: the degenerate surface in the
reduced phase space is the GH restricted to the constraint surface. It is
important to observe that, although in the FLPR model the Gribov ambiguity can
be circumvented by choosing $\lambda=0$ (leading to the analog of the axial
gauge in field theory), an analogous choice is not possible for Yang-Mills
theories. In fact, as shown in \cite{Singer}, in order to include relevant
non-trivial configurations --like instantons-- in the function space of the
theory, certain boundary conditions must be imposed on the fields, which rule
out algebraic gauge conditions (see also \cite{DeW03}). In this sense, a
consistent analog of the limit $\lambda\rightarrow0$ for field theories does
not exist, the Gribov ambiguity is unavoidable for gauge theories, and
degeneracies should be expected in the gauge-fixed system.

As we have shown, when the requirement of regularity is not imposed, a
non-degenerate gauge-fixed systems can be obtained. However this is not a
solution to the problem. Regularity is a key requirement for a set of
constraints to be well defined, as irregularities lead to a Lagrangian that
does not describe the real dynamics of the original system.

Even if the generalization of our results to field theories seems conceptually
straightforward, an interesting future direction for this work is to look for
explicit degeneracies in the gauge-fixed symplectic form in theories of
Yang-Mills type. The problem involves important technical difficulties in the
definition of the reduced phase space when non-algebraic gauge conditions are
adopted. In particular, when set strongly to zero, this kind of gauge
conditions does not allow to express one field as local functions of the
remaining ones, and a local action for the physical degrees of freedom with
the reduced symplectic form is not available. These difficulties in the standard Hamiltonian formulation for Yang-Mills theories make the path integral formalism better suited. However, an interesting novel Hamiltonian approach to QCD, where Dirac reduction is considered, has been recently developed in \cite{Pavel}, which could be worth to study within this context.

The fact that the GH is a degeneracy surface for the gauge fixed system, which
persists at the quantum level, strongly supports the consistency of the Gribov
restriction for QCD, as the degeneracy divides phase space into causally
disconnected regions. Even though the Gribov-Zwanziger idea is heuristic and
supported by the fact that every orbit intersects the Gribov region
\cite{DZ89} (which means that no physical information is lost if the
restriction is applied), the results it yields have gained acceptance by their
match with the lattice data. Our results provide a novel point of view for the
problem in support of the Gribov-Zwanziger proposal that makes it worth to be
studied deeper within the Hamiltonian framework.

\section{Acknowledgments}

We thank M. Astorino, H. Gonz\'{a}lez, O. Miskovi\'{c}, J. Saavedra and A. Toloza for many
enlightening comments and useful discussions. This work has been partially
funded by Fondecyt grants 1140155 and 1120352. The Centro de Estudios
Cient\'{\i}ficos (CECs) is funded by the Chilean Government through the
Centers of Excellence Base Financing Program of CONICYT. P. S-R. is supported
by grants from BECAS CHILE, Comisi\'{o}n Nacional de Investigaci\'{o}n
Cient\'{\i}fica y Tecnol\'{o}gica CONICYT. Partial support by Universidad de
Concepci\'{o}n, Chile is also acknowledged. 

\end{document}